\documentstyle[multicol,prl,aps,epsfig]{revtex}
\newcommand{\be}{\begin{equation}}
\newcommand{\ee}{\end{equation}}
\newcommand{\ba}{\begin{eqnarray}}
\newcommand{\ea}{\end{eqnarray}}
\newcommand{\baa}{\begin{eqnarray*}}
\newcommand{\eaa}{\end{eqnarray*}}
\newcommand{\bea}{\begin{eqnarray*}}
\newcommand{\eea}{\end{eqnarray*}}
\newcommand{\bm}[1]{\mbox{\boldmath $#1$}}
\newcommand{\bb}{\begin{thebibliography}}
\newcommand{\eb}{\end{thebibliography}}

\newcommand{\bi}[1]{\bibitem{#1}}
\begin {document}

\title{Double-spin $\cos\phi$ Asymmetry 
in Semi-inclusive Electroproduction
} 

\author{
K.~A.~Oganessyan,$^{a,b}$
P.~J.~Mulders,$^{c}$
E.~De~Sanctis,$^a$  
} 
%\footnote{e-mail: kogan@mail.desy.de} \\
\address{
$^a$ INFN-Laboratori Nazionali di Frascati I-00044 Frascati, 
via Enrico Fermi 40, Italy \\
$^b$ DESY, Deutsches Elektronen Synchrotron 
Notkestrasse 85, 22603 Hamburg, Germany \\ 
$^c$ Division of Physics and Astronomy, Vrije Universiteit De Boelelaan 1081, \\
NL-1081 HV Amsterdam, the Netherlands  
}

 \maketitle
\begin{abstract}
We consider the double-spin $\cos\phi$ asymmetry for pion electroproduction 
in semi-inclusive deep inelastic scattering of longitudinally polarized 
leptons off longitudinally polarized protons. We estimate the size of the 
asymmetry in the approximation where all twist-3 interaction-dependent 
distribution and fragmentation functions are set to zero. In that 
approximation at HERMES kinematics  a sizable negative $\cos\phi$ 
double-spin asymmetry for $\pi^{+}$ electroproduction is predicted.
\end{abstract}

\begin{multicols}{2}[]
\section{Introduction}

Semi-inclusive deep inelastic scattering (SIDIS) of leptons off a nucleon  
is a suitable process to  extract information  on the quark-gluon
structure -- correlations between  the spins of hadron or quark/gluon 
and the  momentum of the quark/gluon  with respect to  that of  the hadron. 
The parton intrinsic transverse momentum allows in the SIDIS cross 
section particular non-perturbative correlations, which can be probed in 
measurements of azimuthal asymmetries. The complete tree-level result for 
the SIDIS cross section in terms of distribution (DF) and fragmentation (FF) 
functions at leading and subleading order in $1/Q$ has been given in 
Ref.~\cite{TM}. In particular the combination with T-odd fragmentation 
functions leads to single-spin asymmetries. The HERMES collaboration has 
recently reported on the measurement of such single target-spin asymmetries 
in the distribution of the azimuthal angle $\phi$ of produced pions
relative to the lepton
scattering plane, in semi-inclusive charged and neutral pion production 
on a longitudinally polarized hydrogen target~\cite{HERM,HERM1}. 

Other consequences of non-zero intrinsic transverse momentum of partons are
the spin-independent $\cos\phi$ and $\cos2\phi$ asymmetries~\cite{CAHN} 
and double-spin azimuthal asymmetries~\cite{TM,AK}. 
In this paper we investigate the specific $\cos\phi$ azimuthal asymmetry 
at order $1/Q$ in semi-inclusive $\pi^{+}$ production with longitudinally 
polarized electrons on longitudinally polarized protons. 

\begin{figure}[htb]
%\begin{minipage}[t]{75mm}
\epsfxsize 7.5 cm {\epsfbox{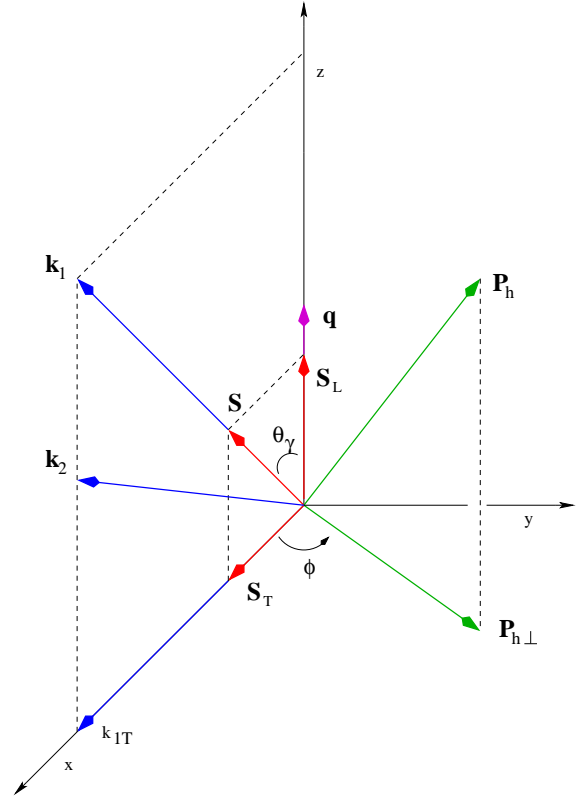}}
\caption{
The kinematics of SIDIS. }
\label{az}
%\end{minipage}
\end{figure}

The kinematics of SIDIS is illustrated in 
Fig.~\ref{az}: $k_1$ ($k_2$) is the 4-momentum of the 
incoming (outgoing) charged lepton, $Q^2=-q^2$, where $q=k_1-k_2$, 
is the 4-momentum of the virtual photon. The momentum
$P$ ($P_h$) is the momentum of the target (observed) hadron. The scaling
variables are $x=Q^2/2(P\cdot q)$, 
$y=(P\cdot q)/(P\cdot k_1)$, and $z=(P\cdot P_h)/(P\cdot q)$. 
The momentum $k_{1T}$ is the incoming lepton
transverse momentum with respect to the virtual photon momentum direction, 
and $\phi$ is the azimuthal angle between $P_{h\perp}$ and $k_{1T}$.
We will consider the case of a polarized beam, the helicity being
denoted by $\lambda_e$. Note that for the specific case in which the target 
is polarized parallel (anti-parallel) to the beam a transverse spin in 
the virtual photon frame arises which only can have azimuthal angle 
$0$ ($\pi$).  The value of this transverse spin component is~\cite{OABK}
\be
\label{TG}
 \vert S_T \vert = \vert S \vert \sin{\theta}_{\gamma}, 
\ee
where $\theta_{\gamma}$ is the virtual photon
emission angle and $S$ is target polarization parallel/antiparallel
to the incoming lepton momentum. 
The quantity $\sin\theta_{\gamma}$ is of
order $1/Q$ and given by
\be
\label{ST}
\sin\theta_{\gamma} = 
\sqrt{\frac{4M^2x^2}{Q^2+4M^2x^2} (1-y-{M^2x^2 y^2 \over Q^2})},
\ee
where $M$ is the nucleon mass. We will distinguish the situations by
referring to $(LL)_{\rm lab}$ which will produce both $LL$ and $LT$
polarization in the virtual photon frame. 
  
\section{The Semi-Inclusive Cross Section}

The cross section for one-particle inclusive deep inelastic 
scattering is given by 
\begin{equation}
\frac{d\sigma^{\ell+N\rightarrow \ell^\prime+h+X}}
{dxdydz d^2P_{h\perp}}=\frac{\pi\alpha^2 y}{2Q^4 z} L_{\mu\nu} 2M 
{\cal W}^{\mu\nu}.  
\label{CS}
\end{equation}

The quantity $L_{\mu\nu}$ is the well-known lepton tensor. The full 
expression for the symmetric and antisymmetric parts of the hadronic 
tensor ${\cal W}^{\mu\nu}$ at leading $1/Q$ order are given by 
Eqs. (77), (78) of Ref. \cite{TM}. In order to investigate 
the {\it cos}$\phi$ azimuthal asymmetry we keep only the terms 
producing contributions in the cross section\footnote{To avoid ambiguities,
we will use the same notations as in Ref. \cite{TM}. }
that are $\phi$-independent or proportional to $\cos\phi$ 
\begin{eqnarray}
2M\,{\cal W}^{\mu\nu} & = &
2 z \int d^2 k_T d^2 p_T
\delta^2(\bm p_T -\bm {P_{h\perp} \over z}- \bm k_T)\nonumber\\ & &
\mbox{}\ \ \times 
\Biggl\{
-g_\perp^{\mu \nu} f_1 D_1+i\,\epsilon_\perp^{\mu \nu}\,g_{1s} D_1
\nonumber\\ & & \quad
+ \frac{2\,\hat{t}^{\{\mu} k_\perp^{\nu \}}}{Q} f_{1} 
\tilde{D}^{\perp}
\nonumber\\ & & \quad
+ \frac{2\,\hat{t}^{\{\mu} p_\perp^{\nu \}}}{Q} x f^{\perp} D_1 
\nonumber\\ & & \quad
+ i\frac{2\,\hat{t}^{\, [\mu} \epsilon_\perp^{\nu ]} k_{\perp \rho}}{Q} g_{1s} 
\tilde{D}^{\perp} 
\nonumber \\ & & \quad 
+ i\frac{2\,\hat{t}^{[\mu} \epsilon_\perp^{\nu]}p_{\perp \rho}}{Q} 
\left[ x g_L^{\perp} D_1 + {M_h \over M} h_{1s}^{\perp} \tilde{E} \right] 
\Biggr\}
\label{WMN}
\end{eqnarray}
where $\{\mu\,\nu\}$ indicates symmetrization of indices and $[\mu\,\nu]$ 
indicates anti-symmetrization.  
In the above expression we have used the 
shorthand notation $g_{1s}$  
\begin{eqnarray}
g_{1s} (x, \bm p_T)= \left[ S_L g_{1L}(x , \bm p_T^2)
+ g_{1T}(x , \bm p_T^2)\,\frac{(\bm p_T \cdot \bm S_T)}{M} \right]  
\label{long}
\end{eqnarray}
and similarly for $h^{\perp}_{1s}$.    

The contraction of leptonic and hadronic tensors leads to the cross section 
with the following terms
\begin{equation}
\frac{d\sigma^{\ell+N\rightarrow \ell^\prime+h+X}}
{dxdydz d^2P_{h\perp}} = \frac{\pi \alpha^2}{Q^2 y} \sum_q e_q^2 
\sigma^q,
\end{equation}
where
\begin{eqnarray}
\sigma^q &=& \int d^2p_T\,d^2\, k_T\, 
z^2 \delta^2(\bm P_{h\perp} - z (\bm p_T - \bm k_T)) 
\nonumber \\ && \mbox{} 
\times \Biggl\{
2\left[1+(1-y)^2\right]\,f^q_1(x,\bm p_T^2)D^q_1(z,\,z^2 \bm k_T^2)
\nonumber \\ && \mbox{}
- 8\,(2-y)\, \sqrt{1-y} \,{1 \over Q}\, \Biggl[ k_{Tx}\,
f_1^q(x,\,\bm p_T^2) \tilde{D}^{\perp\,q}(z,\, z^2 \bm k_T^2) 
\nonumber \\ && \mbox{}
+  p_{Tx}\,x\,f^{\perp\, q}(x,\,\bm p_T^2) D_1^q(z,\, z^2 \bm k_T^2) \Biggr]
\nonumber \\ && \mbox{} 
+2\lambda_e S_L \,y(2-y)
\,g_1^q(x,\,\bm p_T^2)D_1^q(z,\, z^2 \bm k_T^2)
\nonumber \\ && \mbox{}
- 8 \lambda_e S_L\, y \sqrt{1-y} \,{1 \over Q}\, \Biggl[ k_{Tx}\,
g_{1L}^q(x,\,\bm p_T^2) \tilde{D}^{\perp q}(z,\, z^2 \bm k_T^2) 
\nonumber \\ && \mbox{}
+ p_{Tx} \Biggl( x g_L^{\perp q}(x,\,\bm p_T^2) D_1^q(z,\, z^2 \bm k_T^2) 
\nonumber \\ && \mbox{}
+ {M_h \over M } h_{1L}^{\perp q}(x,\,\bm p_T^2) \tilde{E}^q(z,\, z^2 \bm k_T^2) 
\Biggr) \Biggr]
\nonumber \\ && \mbox{} 
+ 2\, \lambda_e\, y(2-y)\, { {(\bm p_T \cdot \bm S_T)} \over M}  
g_{1T}^q(x,\,\bm p_T^2) D_1^q(z,\, z^2 \bm k_T^2).  
\Biggr\} \label{SCS}
\end{eqnarray}
Here by $k_{Tx}$ ($p_{Tx}$) we denote the $x$ component of the 
final (initial) parton transverse momentum vector.
  
\section{The correlation functions}

In the asymmetries considered in this paper a number of functions appear
beyond the well-known leading twist
DF's $f_1^q$, $g_1^q$ and $h_1^q$ and the FF $D_1^q$.
Note that we do not consider polarization in the fragmentation process.
These additional functions are
\begin{itemize}
\item
The DF's $g_{1T}^q$ and $h_{1L}^{\perp\,q}$, interpreted as distributions of
longitudinally and transversely polarized quarks (of flavor $q$) in
transversely and longitudinally polarized nucleons, respectively. The most
interesting $p_T$-integrated functions in these case are the transverse
moments
\be
g_{1T}^{q\,(n)} (x) \equiv \int d^2 p_T\ \left(\frac{\bm p_T^2}{2M^2}\right)^n
\,g_{1T}^q(x,\bm p_T^2).
\ee
\item
The DF's $f^{\perp\,q}$ and $g_L^q$ appear at subleading ($1/Q$) order
in the above expression.
\item
The FF's $D^{\perp\,q}$ and $E^{q}$ appear at subleading ($1/Q$) order. 
\end{itemize}
For all of these functions the relevant transverse moments can be
expressed into the leading twist functions $f_1$, $g_1$, and $h_1$ and
interaction-dependent functions, indicated with a tilde. The relations
are of the same type as the Wandzura-Wilczek relation~\cite{WW} for the 
subleading function $g_T^q$ measured in inclusive leptoproduction with 
a transversely polarized target.

The relations needed in our case and details on them
are found in Refs.~\cite{TM,BHM}. For DF one needs
\ba
\frac{g_{1T}^{(1)}(x)}{x} & = &
\int_x^1 dy\ \frac{g_1(y)}{y}
- \frac{m}{M}\int_x^1 dy\ \frac{h_1(y)}{y^2}
\nonumber \\ &&
- \int_x^1 dy\ \frac{\tilde g_T(y)}{y},
\\
\frac{h_{1L}^{\perp(1)}(x)}{x^2} & = &
-\int_x^1 dy\ \frac{h_1(y)}{y^2}
+ \frac{m}{M}\int_x^1 dy\ \frac{g_1(y)}{y^3}
\nonumber \\ &&
+ \int_x^1 dy\ \frac{\tilde h_L(y)}{y^2},
\\
f^\perp(x) & = & \frac{f_1(x)}{x} + \tilde f^\perp,
\\
g_L^\perp(x) & = & \frac{g_1(x)}{x} + \frac{m}{M}\,\frac{h_{1L}^\perp(x)}{x}
+ \tilde g_L^\perp,
\ea
while for DF we need
\ba
E(z) & = & {m \over M_h}\,z\,D_1(z) + \tilde E(z),
\\
D^{\perp}(z) & = & z\,D_1(z) + \tilde{D^{\perp}}(z).
\ea
The approximation that we will use below consists in setting all interaction
dependent (tilde) functions to zero. There is in fact no justification for
this, except the observation that the same approximation for $g_T^q$
(the Wandzura-Wilczek approximation (WW)) seems to work well~\cite{SLAC}.
We want to add another point concerning potential contributions
in the cross section that have already been neglected, namely those
proportional to $\alpha_s(Q^2)$. Contributions proportional to
$\alpha_s(Q^2) \ldots f_1 \ldots$ will likely appear at the same point where
the function $(M/Q)\ldots f^\perp$ appears~\cite{PG}, while contributions 
proportional to $\alpha_s(Q^2) \ldots g_1 \ldots$ will likely appear at the 
same point where the function $(M/Q)\ldots g_L^\perp$ appears.

\section{Weighted cross section}
 
We will consider the differential cross section integrated over the 
transverse momentum of the produced hadron with different weights 
and denote them by~\cite{KM,BM}
\be
{\langle W \rangle}_{AB} 
= \int d^2P_{h\perp}\, W \frac{d\sigma^{\ell+N\rightarrow \ell^\prime+h+X}}
{dxdydz d^2P_{h\perp}},
\ee
where $W = W(P_{h\perp},\phi,\phi_S)$. With the subscripts $AB$ we denote
the polarization of lepton and target hadron, respectively. 
We use $U$ for unpolarized, $L$ for longitudinally polarized 
and $T$ for transversely polarized particles. 
From Eq.(\ref{SCS}) we then obtain a number of asymmetries. For each of
them we have indicated the results after setting all interaction dependent
functions equal to zero, i.e. only keeping the twist-2 functions. The
results are
\begin{eqnarray}
\label{W1}
\sigma^1_{UU}
& \equiv & {\langle 1 \rangle}_{UU} = \frac{[1+(1-y)^2]}{y}\,f_1(x) D_1(z), 
\end{eqnarray}
\begin{eqnarray}
\label{W2}
\Delta \sigma^1_{LL}
& \equiv & {\langle 1 \rangle}_{LL} = \lambda_e S_L\,(2-y)\,g_1(x)D_1(z),
\end{eqnarray}
\begin{eqnarray}
\label{WUU}
\sigma^2_{UU} & \equiv &
{\langle \vert P_{h\perp} \vert \cos\phi \rangle}_{UU}
\nonumber \\
& = &  - {4 \over Q}\, \frac{(2-y) \sqrt{1-y}}{y} \,   
\Biggl [ M^2 x f^{\perp \, (1)}(x) z D_1(z) 
\nonumber \\ &&\mbox{} \hspace{2cm} - M_h^2 f_1(x) 
z \tilde{D}^{\perp \, (1)}(z) \Biggr] 
\\
& \stackrel{\mathrm{WW}}{\Rightarrow} & 
- {4 \over Q}\, \frac{(2-y) \sqrt{1-y}}{y}\, M^2 f_1^{(1)}(x) z D_1(z),
\label{RU} 
\end{eqnarray}
\begin{eqnarray}
\label{WLL}
\Delta \sigma^2_{LL} &\equiv& {\langle \vert P_{h\perp} \vert \cos\phi \rangle}_{LL}
\nonumber \\
& = & 4 \lambda_e {S_L \over Q}\, \sqrt{1-y}\,  
\Biggl [ M_h^2 g_1(x) z \tilde{D}^{\perp (1)}(z) 
\nonumber \\ 
&&\mbox{}\hspace{2cm} - M^2 x g_L^{\perp (1)}(x) z D_1(z)  
\nonumber \\
&&\mbox{}\hspace{2cm}
- M_h M h_{1L}^{\perp (1)}(x) z \tilde{E}(z)  \Biggr ] 
\\ 
& \stackrel{\mathrm{WW}}{\Rightarrow}  &
- 4 \lambda_e {S_L \over Q}\, \sqrt{1-y}\,   
M^2 g_1^{(1)}(x) z D_1(z) ,
\label{RL}
\end{eqnarray}
\begin{eqnarray}
\label{WLT}
d \sigma^3_{LT} &\equiv &
{\langle \vert P_{h\perp} \vert \cos(\phi-\phi_S) \rangle}_{LT}
\nonumber \\
& = &\lambda_e \vert S_T \vert \, (2-y)\, M\, g_{1T}^{(1)}(x)\, z\, D_1(z)
\\
\label{WLT1}
& \stackrel{\mathrm{WW}}{\Rightarrow} & 
\lambda_e \vert S_T \vert \, (2-y)\, M\,  
\left[\int_x^1 dy\ \frac{g_1(y)}{y}\right]
z\, D_1(z). 
\end{eqnarray} 

The particular $\cos\phi$ moment in the SIDIS cross section for which
we will give an estimate is the following
weighted integral of a cross section asymmetry,

\end{multicols}

\be
{\langle \vert P_{h\perp} \vert \cos\phi \rangle}_{{(LL)}_{lab}} = 
\frac{\int d^2P_{h\perp} {\vert P_{h\perp}\vert}
\cos \phi \left(\sigma^{++}+\sigma^{--}-\sigma^{+-}-\sigma^{-+} \right)}
{{1 \over 4}\int d^2P_{h\perp} \left(\sigma^{++}+\sigma^{--}+
\sigma^{+-}+\sigma^{-+} \right)}.
\label{ASMY}
\ee

%------------------------------------------------------------------------------%
\begin{multicols}{2}[]

Here $\sigma^{++},\sigma^{--} (\sigma^{+-},\sigma^{-+})$ denote the
cross section with antiparallel (parallel) polarization of the 
beam and target respectively\footnote{This leads to positive $g_1(x)$. }. 
They are given by $d\sigma^3_{LT}$ with $\phi_S = \pi (0)$ for $\sigma^{++}$ 
and $\sigma^{-+}$ ($\sigma^{--}$ and $\sigma^{+-})$, respectively.  
The quantity $M_h$ is the mass of the final hadron. Using the 
Eqs. (\ref{W1})--(\ref{WLT1}) and assuming $100 \%$ beam and target 
polarization one obtains  
\be
{\langle \vert P_{h\perp} \vert \cos\phi \rangle}_{{(LL)}_{lab}} =
4 \,\frac{\Delta\sigma^2_{LL}
- d \sigma^3_{LT}}{\sigma^1_{UU}}.  
\label{AS}
\ee
For the experimentally measured cross sections that is determined without
weighing with the transverse momentum of the produced hadron we use 
a further approximation,
\be
A^{\cos\phi}_{{(LL)}_{lab}} \approx {1 \over {\langle P_{h\perp} \rangle} } 
{\langle \vert P_{h\perp} \vert \cos\phi \rangle}_{{(LL)}_{lab}}. 
\label{EXP}
\ee

For the numerical estimate of $A^{\cos\phi}_{{(LL)}_{lab}}$ asymmetry
we use the approximation, where only the twist-2 distribution and fragmentation
functions are used, i.e. the interaction-dependent twist-3 parts
are set to zero.
It is important to point out that in this approximation the $\cos\phi$
asymmetry reduces to a kinematical effect conditioned by intrinsic
transverse momentum of partons similar to the $\cos\phi$ asymmetry in
unpolarized SIDIS~\cite{CAHN}.

\begin{figure}[htb]
\epsfxsize 8.0 cm {\epsfbox{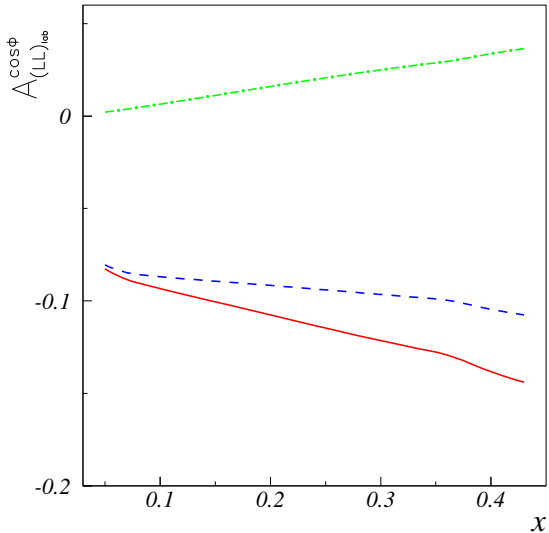}}
\caption{
$A^{\cos\phi}_{{(LL)}_{lab}}$ for $\pi^{+}$
production as a function of Bjorken $x$. The dashed line corresponds to 
contribution of the $\Delta \sigma^2_{LL}$, dot-dashed one to 
$d \sigma^3_{LT}$ and the solid line is the difference of those two. 
}
\label{f2}
\end{figure}

Assuming a Gaussian parameterization for the distribution of the initial
parton's intrinsic transverse momentum, $p_T$, in the helicity distribution 
function $g_1(z,p^2_T)$ one can get 
\be
g_1^{(1)}(x) = \frac{<p_T^2>}{2M^2} g_1(x). 
\label{g1LL}
\ee
It is worth to note that these approximations lead to similar results obtained 
in the simple quark-gluon model with non-zero intrinsic transverse momentum in 
polarized SIDIS~\cite{AK}.  
To estimate the transverse asymmetry contribution $d \sigma^3_{LT}$ into 
the $A_{{(LL)}_{lab}}^{\cos \phi}$, we proceed 
in the same way as in the Ref.~\cite{KM0}.  

In Fig.~\ref{f2}, the asymmetry $A^{\cos\phi}_{{(LL)}_{lab}}(x)$ of Eq.(\ref{EXP}) 
for $\pi^{+}$ production on a proton target is presented as a function of
$x$-Bjorken. The curves are calculated by integrating over the HERMES kinematic 
ranges corresponding to $1$ GeV$^2$ $\leq Q^2 \leq 15$ GeV$^2$, 
$4.5$ GeV $\leq E_{\pi} \leq 13.5$ GeV, $0.2 \leq z \leq 0.7$, 
$0.2 \leq y \leq 0.8$, and taking $\langle P_{h\perp} \rangle = 0.365$ GeV as input. 
The latter value is obtained in this kinematic region assuming a Gaussian 
parameterization of the distribution and fragmentation functions 
with $\langle p_T^2 \rangle=(0.44)^2$ GeV$^2$~\cite{PYTHIA}. 
For the sake of simplicity, $Q^2$-independent parameterizations were chosen  
for the distribution, $g_1(x)$~\cite{BBS}, and fragmentation, $D_1(z)$~\cite{REYA}, 
functions. 

From Fig.~\ref{f2} one can see that the approximation where all twist-3 DF's 
and FF's are set to zero gives the large negative double-spin $\cos\phi$ 
asymmetry at HERMES energies. The `kinematic` contribution to 
$A^{\cos\phi}_{{(LL)}_{lab}}(x)$ coming from the transverse component of 
the target polarization is small (up to $25 \%$ at large $x$-Bjorken). 
 
\section{Conclusion} 
The $\cos\phi$ double-spin asymmetry of SIDIS of 
longitudinally polarized electrons off longitudinally polarized protons 
was investigated. We only kept the $(1/Q)$-order contribution 
to the spin asymmetry that arises from 
intrinsic transverse momentum effects related to twist-two DF and FF   
similar to the $\cos\phi$ asymmetry in unpolarized SIDIS. 
With that approximation, a sizable negative $\cos\phi$ asymmetry is found
for HERMES kinematics.
It is shown that the `kinematical' contribution from target transverse 
component ($S_{T}$) is small. The approximation used to estimate the 
double-spin $\cos\phi$ asymmetry is not complete in $1/Q$ order: it 
contains only $1/Q$ `kinematical' twist-3 contribution. It is similar 
to Cahn's approach~\cite{CAHN} in unpolarized SIDIS, which describes 
well the experimental results from EMC~\cite{EMC} and E665~\cite{E665}. 
The complete behavior of azimuthal distributions needs
the inclusion of higher-twist and pQCD contributions. 
Nevertheless, if one consider the kinematics with $P_{h\perp} < 1$GeV 
and $z < 0.8$, the estimate shows the non-perturbative effects from the 
intrinsic transverse momentum of the partons in the nucleon. The double-spin 
$\cos\phi$ asymmetry is a good observable to investigate the importance 
of leading and sub-leading effects at moderate $Q^2$. 

\vspace*{0.5cm} 

We would like to thank M.~Anselmino, R.~L.~Jaffe and  A.~Sch\"afer for useful 
discussions. This work is part of the research performed 
under the European Commission IHP program under contract HPRN-CT-2000-00130.

\bb{99}
  
  \bi{TM} P.J.~Mulders and R.D.~Tangerman, Nucl. Phys. B {\bf 461} (1996) 197
  and Nucl. Phys. B {\bf 484} (1997) 538, Erratum.
  \bi{HERM} HERMES Collaboration, A.~Airapetian, et al., Phys. Rev. Lett. 
        {\bf 84} (2000) 4047. 
  \bi{HERM1} HERMES Collaboration, A.~Airapetian, et al., Phys. Rev. D {\bf 64} (2001) 097101.
  \bi{CAHN} R.N. Cahn, Phys. Lett.  B {\bf 78} (1978) 269; Phys. Rev. D 
        {\bf 40} (1989)
  \bi{AK} A.~Kotzinian, Nucl. Phys. B {\bf 441} (1995) 234. 
  \bi{OABK} K.~A.~Oganessyan, et al., hep-ph/9808368;  Proc. of the workshop 
            Baryons'98, Bonn, Sept. 22-26, 1998.
  \bi{WW} S. Wandzura and F. Wilczek, Phys. Lett. {\bf B72}, 195 (1977).
  \bi{BHM} D.~Boer, A.~Henneman, P.J.~Mulders, hep-ph/0104271.
  \bi{SLAC} P. L. Anthony,et al., Phys. Lett. B {\bf 458} (1999) 529; 
            K. Abe et al., Phys. Rev. Lett. {\bf 74} (1995) 346.  
  \bi{PG} H.~Georgi, H.D.~Politzer, Phys. Rev. Lett. {\bf 40} (1978) 3.
  \bi{KM} A.M.~Kotzinian, P.J.~Mulders, Phys. Lett. {\bf B406} (1997) 373.
  \bi{BM} D.~Boer and P.J.~Mulders, Phys. Rev. D {\bf 57} (1998) 5780.
  \bi{KM0} A.M.~Kotzinian, P.J.~Mulders, Phys. Rev. D {\bf 54} (1996) 1229.  
  \bi{PYTHIA} T.~Sjostrand, Comp. Phys. Commun. {\bf 82} (1994) 74; CERN-TH.7112/93; 
               hep-ph/9508391. 
  \bi{BBS} S.~Brodsky, M.~Burkardt, I.~Schmidt, Nucl. Phys. {\bf B441} (1995) 197. 
  \bi{REYA} E.~Reya, Phys. Rep. {\bf 69} (1981) 195.
  \bi{EMC} M.~Arneodo, et al., (EMC Collaboration), Z. Phys. C {\bf 34} (1987) 277. 
  \bi{E665} M.~R.~Adams, et al., (Fermilab E665 Collaboration), 
               Phys.Rev. D{\bf 48} (1993) 5057.
\eb

\end{multicols}

\end{document}